\documentclass[10pt,showpacs,superscriptaddress,twocolumn]{revtex4-1}

\usepackage{hyperref}

\usepackage{textcomp,amsmath,amsfonts,amssymb}
\usepackage{graphicx}

\newcommand{\ket}[1]{|#1\rangle}
\newcommand{\bra}[1]{\langle#1|}
\newcommand{\neswarrow}{\mathrel{\text{$\nearrow$\llap{$\swarrow$}}}}

\begin{document}

\title{Fiber-optic realization of anisotropic depolarizing quantum channels}

\author{Micha{\l} Karpi\'{n}ski}
\email{mkarp@fuw.edu.pl}

\author{Czes{\l}aw Radzewicz}
\affiliation{Institute of Experimental Physics, Warsaw University, ul.\ Ho\.z{a} 69, 00-681 Warsaw, Poland}

\author{Konrad Banaszek}
\affiliation{Institute of Physics, Nicolaus Copernicus University,
ul.\ Grudziadzka 5, 87-100 Toru\'{n}, Poland}

\begin{abstract}
We employed an electrically-driven polarization controller to implement anisotropic depolarizing quantum
channels for the polarization state of single photons. The channels were characterized by means of ancilla-assisted
quantum process tomography using polarization-entangled photons generated in the process of spontaneous parametric
down-conversion. The demonstrated depolarization method offers good repeatability, low cost, and compatibility
with fiber-optic setups. It does not perturb the modal structure of single photons, and therefore can be used
to verify experimentally protocols for managing decoherence effects based on multiphoton interference.
\end{abstract}


\maketitle

\section{Introduction}

The practical success of quantum technologies relies on methods to
protect and to enhance quantum effects in the presence of
decoherence. For example, imperfect entanglement between distant
parties can be improved using protocols for entanglement
purification and distillation \cite{Purification}. Transmission and
storage of quantum superpositions can be made more robust with the
help of quantum error correction techniques \cite{QEC}.
Demonstrating experimentally the feasibility of these methods is a
crucial step in the development of quantum technologies.

In this paper we present an experimental realization of a
depolarizing fiber optical channel that can be used to simulate a
range of non-trivial decoherence mechanisms. We anticipate that
the constructed setup can provide a testbed to verify a number of
theoretical proposals for managing decoherence effects. As optical
methods for generating and manipulating entangled states are
relatively well developed, photonic systems offer a promising
route to implement more complex protocols for protecting quantum
information in the presence of noise. Optical approach has been
already used to verify experimentally decoherence-free subspaces
\cite{KwiaBergSCI00,AlteHadlPRL04}, use them in elementary quantum
computing \cite{MohsLundPRL03} and quantum cryptography
\cite{ZhanYinPRA06} protocols, and to carry out single-qubit
purification \cite{RiccDeMaPRL04} and quantum error correction
\cite{PittJacoPRA05}. The device presented in this paper can be programmed
to simulate specific anisotropic depolarizing channels that
appear in practical situations,
for which more efficient specialized protection methods may exist
\cite{LeunVandPRA99,PanSimoNAT01}.

One practical method to generate decoherence in optical qubits is to
introduce correlations between the measured degree of
freedom---typically polarization---and another one, such as the
frequency \cite{KwiaBergSCI00} or the wave vector
\cite{PuenVoigOL06}, that is not sensed by the measuring apparatus,
and from the formal point of view can be considered as the
environment. This approach however modifies the modal structure of
the fields. A polarization qubit prepared initially as a
superposition of one photon in two modes that differ only by their
polarization and are identical otherwise, becomes a complex object
that cannot be assigned a single spatio-temporal profile. This has
profound implications when such qubits are manipulated using presently
the most accessible experimentally tool of multiphoton interference realized in
linear-optics networks. Such
interference requires excellent matching of the spatio-temporal
modes, and its lack provides distinguishing information that
deteriorates interference visibility
\cite{URenBanaszQIC03,HumbGricePRA07}. For demonstrating more complex
manipulations of noise-affected states based on multiphoton
interference, quantum channels in the polarization degree of freedom
need to be implemented without changing the spatio-temporal
properties of the fields, and the approach presented in this paper
satisfies this requirement. In our case, the effective quantum channel is obtained
through a temporal average over suitably chosen polarization transformations.
This approach simulates decoherence rather than induces
it through an interaction with another microscopic ancilla system, with a completely
equivalent outcome.

Our realization of a depolarizing channel is a fiber optic
counterpart of the free-space scheme used in \cite{RiccDeMaPRL04}
based on Pockels cells. Transformations of the state of
polarization (SOP) are introduced with the help of an electrically
driven polarization controller composed of magnetic elements that
introduce birefringence by squeezing mechanically the fiber
\cite{PolControl}. The controller is an inexpensive low-voltage
device, compatible with all-fiber setups and characterized by
negligible insertion losses.
 We verify the action of the channel
through ancilla-assisted quantum process tomography
\cite{AlteBranPRL03} using maximally entangled photon pairs
generated in the process of spontaneous parametric down-conversion
\cite{KwiaPRA99}. Sending one photon through the channel and
performing quantum state tomography on the entire two-photon state
yields complete characterization of the channel thanks to the
Jamio{\l}kowski isomorphism \cite{JamiRMP72}.

We note that scrambling the SOP of optical fields is an important technique
in fiber optic communications, which can reduce the impact of deleterious polarization-dependent
effects in practical transmission systems \cite{YanYuOC05,PerlOC05,LizeGommOC07}. Compared
to applications in high bit rate fiber communication systems, requirements set by
proof-of-principle demonstrations of quantum protocols are somewhat different. Firstly, shortening time scale
of polarization scrambling is not of primary importance, as typical experiments with entangled photons
require long collection times. Therefore mechanical scramblers are presently well suited for this class of
applications. Secondly, the actual distribution of the SOP after scrambling is not important, as only
ensemble averages matter for experiments in which consecutive outputs are manipulated and
detected individually.

The structure of this paper is as follows. In Sec.~\ref{Sec:Theory}
we summarize the theoretical basis of our experiment.
The experimental setup is presented in Sec.~\ref{Sec:ExpSetup}, and obtained
results are discussed in Sec.~\ref{Sec:Results}. Finally, Sec.~\ref{Sec:Concl}
concludes the paper.

\section{Theory} \label{Sec:Theory}

Let us start with a brief theoretical review of qubit channels. The
most convenient description is based on the Bloch vector
representation. A general channel ${\cal E}$ is given by an affine
map acting on the Bloch vector $\mathbf{r}$ according to
\cite{KingRuskTIT01}:
\begin{equation} \label{Eq:Map}
{\mathbf r} \mapsto {\boldsymbol\Lambda} {\mathbf r} + {\mathbf
a},
\end{equation}
where ${\boldsymbol\Lambda}$ is a real $3\times 3$ matrix and
${\mathbf a}$ is a vector characterizing the output state for a
completely mixed input. We will be interested here only in unital
maps, which do not change the maximally mixed state, and therefore
${\mathbf a} = 0$. Singular value decomposition enables us to
represent ${\boldsymbol\Lambda}$ as a product
 ${\boldsymbol\Lambda} = \mathbf{O} \mathbf{D} \mathbf{O}'$ of two proper
rotations
 $\mathbf{O}, \mathbf{O}'$ and
a diagonal matrix $\mathbf{D}=\text{diag}(D_x,D_y,D_z)$. This allows
us to separate the trivial rotations of the Bloch vector
corresponding to unitary transformations from the depolarization
process, described by the three singular values comprising
$\mathbf{D}$. Distinctly from the usual convention we allow here
negative singular values, which stems from the requirement that the
rotations $\mathbf{O}, \mathbf{O}'$ are proper.

\begin{figure}
\begin{center}
    \includegraphics{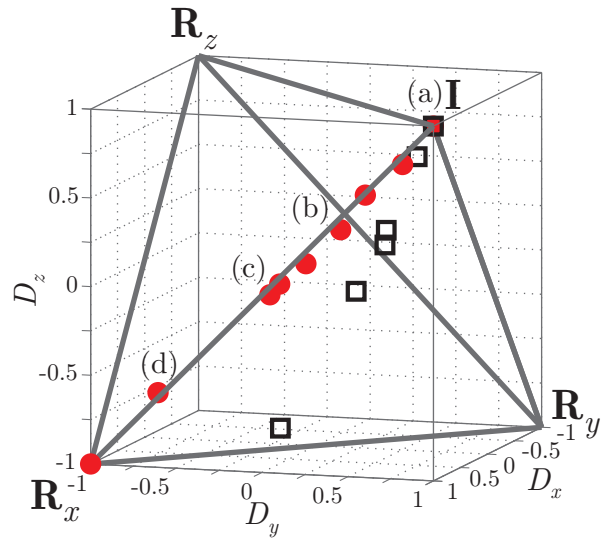}
    \caption{The tetrahedron of singular values for unital quantum channels which
    results from the physical condition of complete positivity. Dots and
    squares represent channels implemented and characterized experimentally
    by means of ancilla-assisted quantum process tomography, discussed in detail
    in Sec.~\protect\ref{Sec:Results}. Roman labels (a-d) refer to channels depicted in
Fig.~\protect\ref{Fig:Edge}.\label{Fig:Tetra}} 
\end{center}
\end{figure}

Physical channels must satisfy the condition of complete positivity,
which means that a trivial extension of the channel ${\cal E}$ to any
auxiliary system yields a positive map. For unital maps this
condition implies the following constraints on $\mathbf{D}$
\cite{KingRuskTIT01}:
\begin{equation}
\label{Eq:TetraCond} |D_x \pm D_y | \le | 1 \pm D_z|,
\end{equation}
where the $\pm$ signs on the left- and the right-hand side of the
above equation are taken independently. Geometrically, the set of
allowed parameters $D_x,D_y,D_z$ forms a tetrahedron shown in
Fig.~\ref{Fig:Tetra}, whose vertices are given by the identity
$\mathbf{I}$ and 180$^\circ$ rotations $\mathbf{R}_x, \mathbf{R}_y,
\mathbf{R}_z$ about axes of the coordinate system in the Bloch
representation. These rotations correspond to unitary
transformations described by Pauli matrices $\hat{\sigma}_x,
\hat{\sigma}_y, \hat{\sigma}_z$. The geometrical picture shows that
the non-unitary depolarizing part of an arbitrary unital channel
given by $\mathbf{D}$ can be represented as a convex combination of
the identity and three rotations:
\begin{equation}
\label{Eq:LambdaDec} \mathbf{D} = p_0 \mathbf{I} + p_x \mathbf{R}_x
+ p_y \mathbf{R}_y + p_z \mathbf{R}_z,
\end{equation}
where the nonnegative parameters $p_0, p_x, p_y, p_z$ form a
probability distribution. The decomposition given in
Eq.~(\ref{Eq:LambdaDec}) defines a practical way to implement an
arbitrary unital channel by a random applications of identity
$\mathbf{I}$ and rotations $\mathbf{R}_x, \mathbf{R}_y,
\mathbf{R}_z$ with appropriate probabilities, sandwiched between
unitary transformations that realize $\mathbf{O}$ and $\mathbf{O}'$.
This approach will be the basis of our experiment.

We will characterize the action of the channel using
ancilla-assisted quantum process tomography, which is based on the
Jamio{\l}kowski isomorphism \cite{JamiRMP72} between quantum
channels and bipartite states. Specifically, preparing two qubits in
a maximally entangled bipartite state $\ket{\Phi}$ and sending one
of the subsystems through the channel yields a state
$\hat{\varrho}_{\cal E} = ({\cal E} \otimes
I)(\ket{\Phi}\bra{\Phi})$, that carries complete information about
the channel. Performing quantum state tomography on
$\hat{\varrho}_{\cal E}$ allows one to characterize fully the
channel ${\cal E}$.  The density matrix of the maximally entangled
input state can be written in the form \cite{HoroHoroPRA96}:
\begin{equation}
\label{Eq:Phi+Phi+} |\Phi\rangle\langle\Phi| = \frac{1}{4}
\left(\hat{\openone} \otimes \hat{\openone} -\!\!\!\!\!\!
\sum_{i,j=x,y,z}\!\!\!\!\! T_{ij}\, \hat{\sigma}_{i} \otimes
\hat{\sigma}_{j}\right),
\end{equation}
where $\mathbf{T}$ is a real $3\times 3$ matrix with the singular
values equal to $\pm 1$. For example, a state $\ket{\Phi_+} =
(\mid\leftrightarrow\leftrightarrow\nolinebreak[4]\rangle +
\mid\updownarrow\updownarrow\nolinebreak[4]\rangle)/\sqrt{2}$, close
to the one used in our experiment, corresponds to $\mathbf{T} =
\text{diag}(-1,1,-1)$. For unital channels, the output state
$\hat{\varrho}_{\cal E}$ has a form analogous to
Eq.~(\ref{Eq:Phi+Phi+}) with the matrix $\mathbf{T}$ replaced by
${\boldsymbol\Lambda}\mathbf{T}$. Its elements can be calculated as
expectation values $({\boldsymbol\Lambda}\mathbf{T})_{ij} =
\text{Tr}[\hat{\varrho}_{\cal E}(\hat{\sigma}_{i} \otimes
\hat{\sigma}_{j})]$. Multiplying the reconstructed matrix by
$\mathbf{T}^{-1}$ yields ${\boldsymbol\Lambda}$. This method
tolerates imperfections in the input state such as mixedness, as
long as the matrix $\mathbf{T}$ is invertible.
 Because our primary
interest here are depolarizing properties of quantum channels and
unitary rotations can be easily compensated  with polarization
controllers, we will use experimental data to extract the singular
values of ${\boldsymbol\Lambda}$ that are responsible for the
non-unitary dynamics.

It is worth noting that applying the depolarization method described
above to a maximally entangled state allows one to generate an
arbitrary state of the form given by the right hand side of
Eq.~(\ref{Eq:Phi+Phi+}). The matrix ${\mathbf T}$ can be brought to
the diagonal form analogously to the Bloch representation of unital
channels, with the singular values satisfying the analog of
Eq.~(\ref{Eq:TetraCond}) which stems from the positivity condition
for density matrices. The output state can be fully controlled through
voltages applied to a single device, unlike sources of mixed
bipartite states which rely on more complex preparation procedures
\cite{BarDeMaPRL04}.

\section{Experimental setup}
\label{Sec:ExpSetup}

\begin{figure}
\begin{center}
   \includegraphics{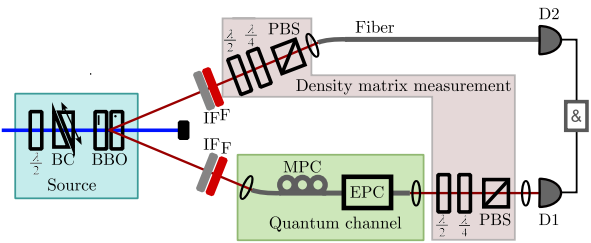}
   \caption{Experimental setup. BC, Babinet compensator; BBO,
   a pair of beta-barium borate crystals; IF, interference filter; F, red filter; MPC, manual
   fiber polarization controller;  EPC, electrically-driven fiber polarization
controller; PBS, polarizing beam splitter;
   D1, D2, single-photon detectors.
  \label{Fig:ExpSetup}} 
\end{center}
\end{figure}

The experimental setup is shown in Fig.~\ref{Fig:ExpSetup}. The
ultraviolet pump beam entering the setup from the left is prepared
by doubling femtosecond pulses from a titanium-sapphire laser (40 fs
FWHM pulse duration, 80 MHz repetition rate, 780~nm central
wavelength) in a beta-barium borate (BBO) crystal to yield second
harmonic with the average power of 14~mW, which is separated from
the fundamental with two dichroic mirrors and a blue color filter.

The source of entangled photon pairs is based on spontaneous
parametric down-conversion in a pair of type-I BBO crystals
oriented at 90$^\circ$ with respect to each other
\cite{KwiaPRA99}. The half-wave plate $\lambda/2$ and the Babinet
compensator BC preceding the crystals ensure that the pairs
produced by both crystals are temporally indistinguishable and
have equal probability amplitudes. The generated signal and idler
photons are sent through red filters F and 5~nm bandwidth
interference filters IF to filter out the scattered pump field and
to restrict the spectral bandwidth.

The idler photons are then coupled into a single-mode fiber using an
aspheric lens. They pass first through a manual fiber polarization
controller MPC with settings fixed at the beginning of the
experiment to compensate for the overall SOP change in the fiber,
and then through the electrically driven polarization controller EPC (OZ-Optics
model EPC-400) which implements the actual depolarization
channel. The idler photons are subsequently coupled out of the fiber
into a collimated beam with an aspheric lens
 and sent through a quarter-, a
half-wave plate and a polarizer to select a specific polarization
state, detected finally with a free-space single-photon counting
module D1 (Perkin Elmer model SPCM-AQ-131). An identical set of
two wave-plates and a polarizer is placed in the path of the
signal photons before coupling them into a single-mode fiber
connected directly to a module D2 (Perkin Elmer model
SPCM-AQR-14-FC). To reconstruct the density matrix, coincidence
count rates were measured in a basis of 16 different two-photon
polarization states that form a tomographically complete set. In
the standard notation, the basis consisted of the following
states: $\ket{\!\!\leftrightarrow\leftrightarrow}$,
$\ket{\!\!\leftrightarrow\updownarrow}$,
$\ket{\!\!\updownarrow\updownarrow}$,
$\ket{\!\!\updownarrow\leftrightarrow}$,
$\ket{\!\!\circlearrowright\leftrightarrow}$,
$\ket{\!\!\circlearrowright\updownarrow}$,
$\ket{\!\!\neswarrow\updownarrow}$,
$\ket{\!\!\neswarrow\leftrightarrow}$,
$\ket{\!\!\neswarrow\circlearrowright}$,
$\ket{\!\!\neswarrow\neswarrow}$,
$\ket{\!\!\circlearrowright\neswarrow}$,
$\ket{\!\!\leftrightarrow\neswarrow}$,
$\ket{\!\!\updownarrow\neswarrow}$,
$\ket{\!\!\updownarrow\circlearrowleft}$,
$\ket{\!\!\leftrightarrow\circlearrowleft}$,
$\ket{\!\!\circlearrowright\circlearrowleft}$, selected by
appropriate settings of quarter- and half-wave plates in both
arms. The tomographic reconstruction of the density matrices was
carried out using the maximum likelihood technique
\cite{BanaszPRA99,JamKwiatPRA01}.

The electrically driven fiber polarization controller consists of
four magnetically driven squeezers driven by external voltages,
which apply mechanical stress to the optical fiber thus inducing
birefringence. Each squeezer works as a wave-plate with a fixed
orientation and a tunable phase delay. It is easy to see that
placing a sequence of three such squeezers with relative
45$^\circ$ rotations between their optical axes  allows one to
implement an arbitrary rotation of the Bloch sphere, with phase
delays defining the three Euler angles of the rotation
\cite{PolControl}. The device used in the experiment included four
squeezing elements, and we found that their relative orientation
required more complex driving to achieve necessary polarization
rotations. Switching between several different settings of applied
voltages and averaging over time yields a depolarizing channel
with controllable characteristics.
The time of averaging was determined by the brightness of our
source of entangled photon pairs and was chosen to be 70~s for
each setting of the quarter- and half-wave plates needed for
density matrix reconstruction. That time was much larger compared
to the duration of a single setting of the polarization
controller, which was of the order of 100 ms (exact figures will
be specified in Sec.\ \ref{Sec:Results}). By employing a
polarization controller faster then the one used in our experiment
\cite{LizeGommOC07}, this depolarization method could be used with
much shorter acquisition times and thus with brighter sources.
 The key advantage of this
depolarization method is that it does not introduce correlations
between the polarization state and spatial or spectral degrees of
freedom of transmitted photons, as long as polarization mode
dispersion can be neglected over the spectral range of the
transmitted photons.

We define the quantum channel whose characteristics are to be
measured as the polarization transformation experienced by the idler
photons between the aspheric lenses coupling into and out of the
fiber. The action of EPC was calibrated by coupling into the channel
a macroscopic laser beam prepared in three different states of
polarization: horizontal, diagonal, or left-circular, and performing
a standard SOP measurement on the output using an ellipsometer that
detected intensities of horizontal-vertical, diagonal, and circular
components on three parts split off the input beam. In the
following, we will refer to this measurement as simple process
tomography to differentiate it from the ancilla-assisted quantum
process tomography, that is the core of our experiment. In the first
step of the calibration procedure, the driving of the EPC was
switched off completely and MPC was aligned manually such that the
action of the channel was equal to the identity. Starting from this
reference state, voltage settings required to achieve 180$^\circ$
rotations were determined.


\section{Results}
\label{Sec:Results}

\begin{figure}
  \centering
     \includegraphics{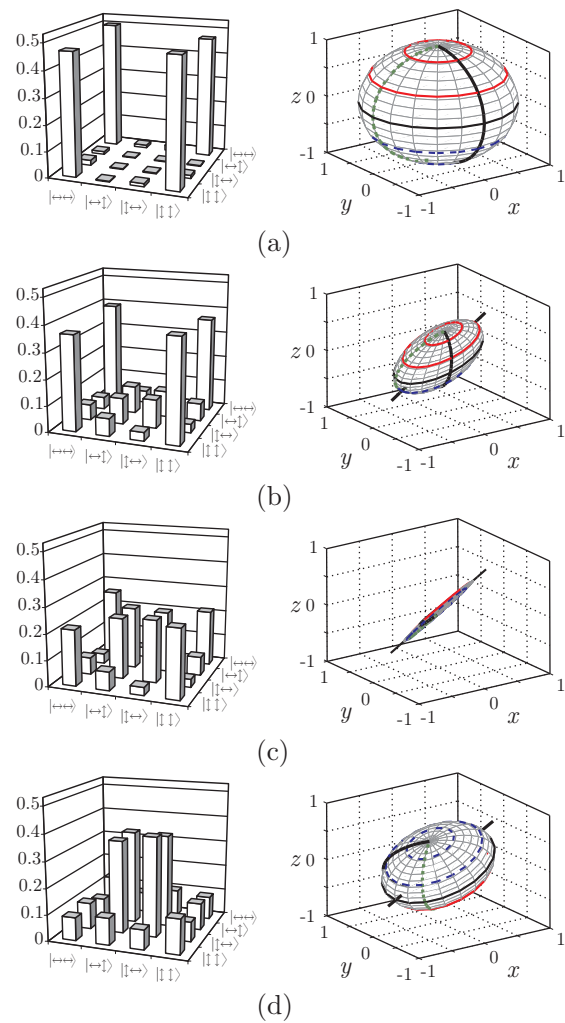}
  \caption{The absolute values of density matrix elements (left) and ellipsoids representing
  the transformed Bloch sphere
  (right) measured by ancilla-assisted quantum process tomography for quantum channels
  obtained by switching between the identity and a selected 180$^\circ$ rotation; (a)
  the input density matrix and the initial Bloch sphere, with selected
  parallels and meridians marked for reference; (b-d) quantum channels obtained
  by changing the duty cycle of the voltage driving the controller.
  Note that the points representing
  the poles of the original sphere do not coincide with the principal axes
  of the ellipsoids, marked with thick straight black lines.
  \label{Fig:Edge}}
\end{figure}

The input state for ancilla-assisted quantum process tomography was
prepared by setting the action of the quantum channel to identity
and measuring the density matrix of the generated state. For
measurements described in this paper, coincidence counts were
collected over 70~s for each pair of polarization settings. The
results of quantum state tomography performed on the input state are
shown in Fig.~\ref{Fig:Edge}(a). The linearized entropy of this
state is equal to $0.1\pm 0.1$, and the amount of entanglement can
be quantified with the concurrence value of $0.92 \pm 0.07$. The
input state is well approximated by a pure state of the form
$\ket{\Phi_+} =
(\mid\leftrightarrow\leftrightarrow\nolinebreak[4]\rangle +
\exp(i\phi)
\mid\updownarrow\updownarrow\nolinebreak[4]\rangle)/\sqrt{2}$
 with the phase equal to $\phi=-0.12$. The fidelity between the
measured state and its pure-state approximation is equal to $0.98
\pm 0.09$.

We concentrated our efforts on the
implementations of anisotropic channels. In the first series of measurements we studied
channels that lie on the edge of the tetrahedron, and can be
realized as a mixture of the identity $\mathbf{I}$ and a single
180$^\circ$ rotation $\mathbf{R}$, which is obtained by applying an
appropriate voltage to one of the squeezers in the EPC. Thus, the
realized quantum channel is given, up to rotations $\mathbf{O}$ and
$\mathbf{O}'$, in the form
  ${\boldsymbol\Lambda} = (1-p)\mathbf{I} + p\mathbf{R}$.
It is easily seen that one of the singular values of such a
transformation is equal to 1 irrespectively of $p$, while the
remaining two are identical and given by $1-2p$. The voltages
driving the EPC were programmed in a temporal loop with a 150~ms
period, and the parameter $p$ defined the duty cycle during which
the EPC operated as $\mathbf{R}$.

The experimental results are shown in Fig.~\ref{Fig:Edge}(b-d). The
left column depicts output density matrices reconstructed using the
maximum likelihood algorithm, while the right column represents
measured quantum channels with the help of ellipsoids that are
obtained by acting with the channels on the input Bloch sphere.
It
is worth noting that the principal axes of all the ellipsoids are
oriented along the same direction to within 3$^\circ$. This is
consistent with our expectations: the transformation $\mathbf{R}$
realized by the EPC is applied along the same direction determined
by the orientation of the mechanical squeezer in the EPC, and the
birefringence of the input and output fibers, which causes rotations
of the SOP, is also fixed.

 By means of simple process tomography we
found that after taking off the voltage implementing $\mathbf{R}$
the EPC did not return to the state corresponding to the identity
channel. We attribute this effect to the residual hysteresis in
the magnetic elements of the EPC.
 We partially compensated it when
implementing channels $(1-p)\mathbf{I} + p\mathbf{R}$ by setting a
small voltage with an opposite sign during time intervals when the
identity transformation $\mathbf{I}$ was to be realized.

\begin{figure}
\centering
   \includegraphics{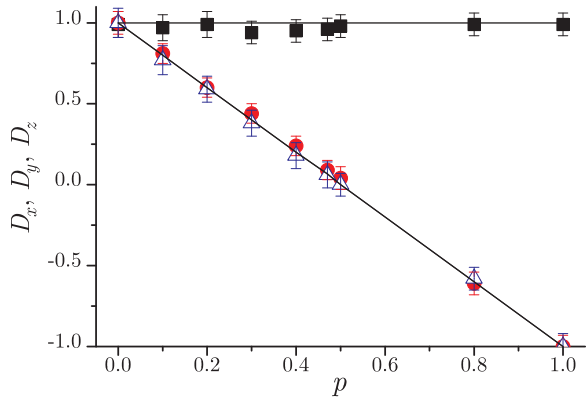}
   \caption{A graph of the singular values for channels
   of the form $(1-p)\mathbf{I} + p\mathbf{R}$ as a function of the
   duty cycle $p$. The three singular values corresponding to an experimentally characterized
   channel are represented by a
   square, a circle, and a triangle. Solid lines depict
   values predicted directly from the value of the duty cycle
   as $(1,1-2p,1-2p)$ \label{Fig:SingVal}}
\end{figure}

The ellipsoids depicted in Fig.~\ref{Fig:Edge} contain complete
information about quantum channels, including unitary
transformations that accompany depolarization. In order to analyze
more closely the depolarizing characteristics of the realized
channels, in Fig.~\ref{Fig:SingVal} we plot the triplets of the
singular values versus the programmed duty cycle $p$ which defines
the share of $\mathbf{R}$ in the channel. A good agreement between
the predicted and measured singular values is clearly seen.

In the second series of measurements, we implemented quantum
channels that are described by mixtures of the identity and two
180$^\circ$ rotations about orthogonal axes, which we will denote as
$\mathbf{R}_1$ and $\mathbf{R}_2$. Such a set of unitary
transformations allows one to produce all of the transformations
represented by points located on a face of the tetrahedron shown in
Fig.~\ref{Fig:Tetra}. Using simple process tomography we searched
for voltage settings that would implement such a pair of rotations.
The best approximation we were able to find required simultaneous
driving of all four squeezers. The angle between the rotation
axes was determined to be 86$^\circ$~$\pm$~3$^\circ$, with the
rotation angles given by 175$^\circ$~$\pm$~3$^\circ$.
These findings strongly suggest that the alignment of the squeezers
inside the EPS must have been significantly different from the
45$^\circ$ relative rotations that give parametrization of the polarization
transformation in terms of the Euler angles.

\begin{figure}
  \centering
  \includegraphics{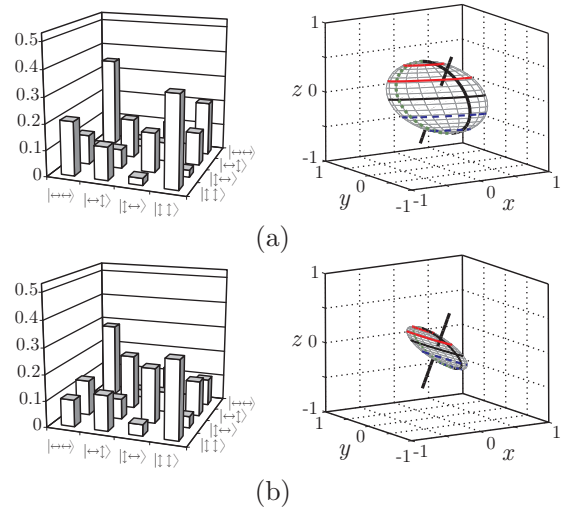}
  \caption{
  The absolute values of density matrix elements (left) and ellipsoids (right), representing
  the transformed Bloch sphere shown in Fig.~\ref{Fig:Edge}(a),
   measured by ancilla-assisted quantum process tomography for quantum channels
  obtained by switching between the identity and two 180$^\circ$ rotations with approximately
  orthogonal axes. Straight thick black lines mark the orientation of the shortest principal
  axis of an ellipsoid. \label{Fig:Face}}
\end{figure}

We used the rotations $\mathbf{R}_1$ and $\mathbf{R}_2$ to realize a
series of maps of the form $\Lambda = (1-p)\mathbf{I} +
p(\mathbf{R}_1 + \mathbf{R}_2)/2$. Such maps are represented by
points located on the median of one of the faces of the tetrahedron
depicted in Fig.~\ref{Fig:Tetra}. Analogously to the first series of
measurements, the settings of the EPC were updated in a temporal
loop with a period of 50~ms. Within a single interval, each of the
rotations $\mathbf{R}_1$ and $\mathbf{R}_2$ was realized during a
fraction $p/2$ of the interval, followed by an application the
identity $\mathbf{I}$, with voltage settings taking into account the
partial compensation for the hysteresis.

\begin{figure}
  \centering
  \includegraphics{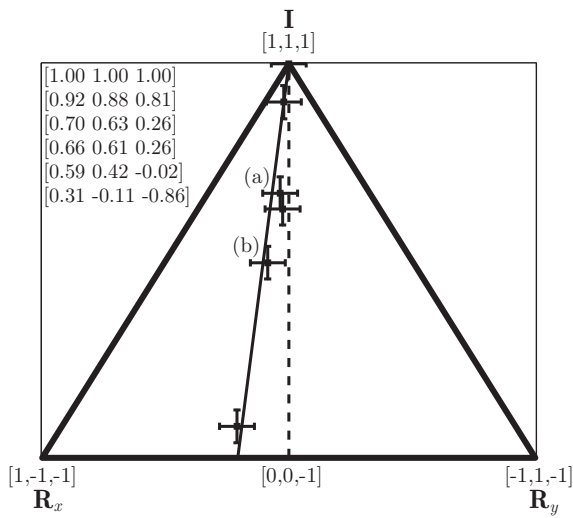}
  \caption{An orthogonal projection of triplets of singular values for experimentally implemented channels
  of the form $(1-p)\mathbf{I} + p(\mathbf{R}_1 + \mathbf{R}_2)/2$ onto the tetrahedron surface
  The solid line is the least square fit to the experimental points,
  compared to the median marked with a dashed line. The points labelled with (a) and (b) correspond to channels
  depicted in Fig.~\protect\ref{Fig:Face}. The singular values are given explicitly in the upper left
  corner of the plot. \label{Fig:Projection}} 
\end{figure}

Exemplary results are presented in Fig.~\ref{Fig:Face} depicting
output density matrices and Bloch sphere transformations. In order
to analyze the results more closely, we extracted the singular
values and projected them onto the tetrahedron face, shown in
Fig.~\ref{Fig:Projection}. Although the rotations $\mathbf{R}_1$
and $\mathbf{R}_2$ were applied over identical fractions of the
cycle, the reconstructed maps do not lie on the median of the
tetrahedron face. This difference cannot be fully explained by the
facts that the rotations $\mathbf{R}_1$ and $\mathbf{R}_2$ were
characterized by angles deviating from 180$^\circ$ and that their
axes were not strictly perpendicular. We attribute the observed
disagreement to the hysteresis in the squeezing elements of the
EPC. Nevertheless, it is worth noting that within the experimental
errors the triplets of the singular values do form a straight line
located on a face of the tetrahedron. The mismatch between the
predicted and the observed characteristics of the channel should
not pose a substantial difficulty in practice, because the
averaged operation of the EPC over the entire period is repeatable
to a very good degree.
 Therefore it is possible to determine, via process
tomography, voltage settings that realize a required quantum
channel, and use them in further experiments.
 Moreover, further
improvement could be made by using a larger number of squeezing
elements, equal to 5 or 6, which, as suggested in \cite{PerlOC05},
should significantly improve the performance of the polarization
controller.

\section{Conclusions}
\label{Sec:Concl}

We used an electrically driven polarization controller to
implement unital quantum channels acting on the polarization state
of a single photon. The characteristics of the channels were
verified with the help of ancilla-assisted quantum process
tomography. Although the controller requires a calibration
procedure for voltage settings to implement a given channel, its
operation is repeatable to a very good degree. Therefore it can be
reliably used in more complex experiments testings methods to
manage decoherence effects in emerging quantum-enhanced
technologies. Its advantages are low cost, uncomplicated driving,
and compatibility with fiber-optic setups.

This work was carried out at the National Laboratory of Atomic,
Molecular and Optical Physics (FAMO) in Toru\'n and was supported by
the European Commission under the Integrated Project Qubit
Applications (QAP) funded by the IST directorate as Contract Number
015848, Polish MNiSZ grant 1~P03B~011~29, and AFOSR under grant
number FA8655-06-1-3062.

\end{document}